# Thermodynamic adsorption potential of superconductors

Jiu Hui Wu [1], Jiamin Niu [1], and Kejiang Zhou [2]

*[1] School of Mechanical Engineering, Xi'an Jiaotong University,*
*& State Key Laboratory for Strength and Vibration of Mechanical Structures, Xi'an 710049, China*

*[2] Huzhou Institute of Zhejiang University, Huzhou 313000, China*

Abstract: Based on the general thermodynamic analysis of Polanyi adsorption potential, the adsorption potential condition for superconductors is obtained exactly by using the quantum state equation we presented. Because this adsorption potential results in changes of electron concentration, temperature and pressure in a certain volume (adsorption space) adjacent to the surface of the lattice, the composition and structure of superconductors are of course decisive for the adsorption potential. Then we calculate the molar adsorption potentials for those typical superconductors, and find that it is positively correlated to the superconductivity temperature $T_c$, which reveals that those high-$T_c$ superconductors are mainly determined by the higher molar adsorption potentials. In addition, the adsorption potential at $T = T_c$ still works despite the disappearance of the energy gap of the BCS theory. This shows that beyond the electron-phonon interaction mechanism, the Cooper-paired electrons are mainly formed by this physical adsorption potential for high-$T_c$ superconductors. This adsorption potential theory could explain almost all common facts about high-temperature superconductors, including many anomalies of the normal and superconducting states.

## 1. Introduction

Since the discovery of superconductivity by Kamerlingh Onnes in 1911, the study of superconductivity has been a hot topic and a difficult point in the field of physics, with a history of more than 100 years. In the early days, researches on superconductors focused on metal and alloy systems with low superconducting transition temperatures, whose mechanism can be explained by the BCS theory[1] founded in 1957, and related theories that have been developed subsequently. Based on the BCS theory, it is difficult for conventional superconductors to have a $T_c$ higher than 40K at atmospheric pressure[2], so superconductors with $T_c$ higher than 40K are often referred to as high-temperature superconductors. In 1986, Müller and Bednorz discovered the $(La,Ba)_2CuO_4$ superconductor[3], and in 1987 the discovery of $YBa_2Cu_3O_{7-\delta}$ raised the superconductivity temperature to the liquid nitrogen temperature region for the first time [4,5], opening a new era of high-temperature superconductivity research. The $T_c$ of these copper-based materials is much higher than that of conventional superconductors, and the highest atmospheric pressure is currently held at 133 K in 1993 [6]. Since 1990s, high-precision epitaxial growth technology has been introduced into superconductivity research, and the heterostructures can be formed by artificially stacking materials with different structures or different chemical compositions [7], and novel superconducting states may also be derived based on the coupling of heterogeneous atomic-level smooth interfaces [8–13]. Through the artificial design of heterostructures or precise control of the thickness of single crystal films, researchers are able to explore superconductivity under dimensional changes in highly crystalline samples [14]. More recently, Yanagisawa [15] investigated correlated-electron systems, emphasizing strong electron correlation as a driver of superconductivity in high-temperature Cuprates.

Up to date, the results of many experiments [16] reveal the following common facts about high-temperature superconductors: 1) They are strongly correlated electronic systems; 2) It is important to consider some kind of electron-electron mediation; 3) Due to their severe anisotropy, the interlayer coupling must be considered; 4) There should be an appropriate concentration of carriers, too much or too little is not good for superconductivity; 5) It is necessary to establish the theory of superconductivity on the basis of the anomalous electronic states of the normal phase, because the anomaly of the normal phase contains the special interaction mechanism of the electronic system.

At present, the mechanism of high-temperature superconductivity is still unsolved, and the journal Science has repeatedly included it in 125 important scientific questions. Due to the strong interaction between electrons in high-temperature superconductors, the description of their electronic behavior inevitably involves the extremely complex many-body quantum physics, so the mechanism of high-temperature superconductivity is still not fully understood, especially the source of the attraction effect used for pairing between carriers is still the core of the current physics field one of the difficulties.

Grounded in the work of Henri Poincare, R. Thom and others [17,18], the catastrophe theory can explain the phenomenon of gradual quantitative change to sudden qualitative change, which is a highly generalized mathematical theory that summarizes the rules of non-equilibrium phase transition by several catastrophe models. According to Thom's classification theorem, as long as the number of control variables that cause mutations does not exceed 4, the various mutation processes in nature can be grasped using seven basic potential function models. Because of its own structural-stability during the mutation process of any non-equilibrium system, any phase transition can be analyzed quantitatively by one of these several catastrophe models, even without knowing the differential equations of the system. In our previous papers, by using the catastrophe theory, the general non-equilibrium phase transition process from laminar to turbulent has been investigated quantitatively [19], as well as a revised Schrödinger relativistic equation obtained from the perspective of phase transition [20].

More recently, the thermodynamic quantum phase transition process was investigated quantitatively by the structural-stability-based catastrophe theory [21]. For a canonical ensemble composed of $N$ identical particles, the cusp catastrophe model is adopted to express the average free energy, and further the general quantum state equation about the pressure is obtained to describe the phase transition process from quantum scale to macro scale by using the dimensionless analysis. Subsequently, the ensemble free energy with considering the interaction potential energy among the particles can be gotten exactly, as well as the canonical partition function and the specific heat capacity of the system.

The transition from a normal state to a superconducting state is a phase transition problem, which obviously also falls under the category of thermodynamics. The superconductivity is a second-order phase transition, that is, the superconducting phase transition does not undergo latent heat changes, and meanwhile the free energy of the superconducting states must be less than that of the normal states. Analogy with Polanyi adsorption potential theory in chemistry [17], the physical adsorption potential between free electrons and the lattice of superconductors through the van der Waals force is put forward, assuming that these electrons can be adsorbed and condensed due to changes in electron concentration, temperature and pressure in a certain volume (adsorption space) adjacent to the surface of the lattice, in addition to the electron-phonon interaction mechanism.

In this paper, based on this general quantum state equation by the catastrophe theory, the thermodynamic adsorption potential of superconductors is put forward to try to explain the high-temperature superconductivity.

## 2. General thermodynamic analysis of Polanyi adsorption potential

The adsorption of gas or liquid on the solid surface is the result of the adsorption displacement caused by the adsorption force between the adsorbate in the adsorption force field and the solid surface. The field strength of an equipotential surface in the adsorption force field is not only related to the properties of the adsorbate and the solid, and its surface structure, but also affected by the distance between the equipotential surface and the solid surface. The field strength is decreased with increase of the distance, and becomes zero at infinity, which is defined as the zero-potential surface.

In the adsorption force field, the adsorbate moves along the direction of the field strength under the action of the adsorption force to do the adsorption work. Here the effect of adsorption force is greater than that of molecular thermal movement, which results in the reduced distance among the adsorbate molecules, and even produces coagulation or chemical adsorption. The adsorption layer between the normal phase of the adsorbate and the surface of the solid is called the adsorption phase, in which the adsorbate density is greater than that of the normal phase, and the concentration of the adsorption phase showed a gradient and continuous change, as shown in Fig.1.

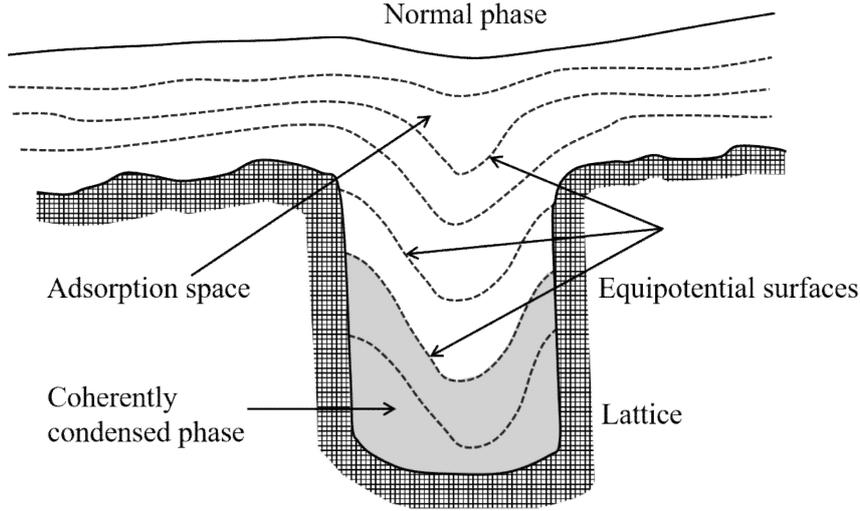

Fig.1 Diagram of the physical adsorption potential adjacent to the surface of the lattice

From the above, it can be seen that the adsorption process is a process in which the adsorbate molecules enter the adsorption phase from the normal phase under certain conditions, and the displacement change or chemical change occurs, that is, the amount of the substance changes. Below the chemical potential and adsorption work are used to reflect the rule of the adsorption process.

For a system consisting of the components of the adsorbate, the adsorption force is a generalized force, and the Gibbs free energy of the adsorption system is expressed as

$$G = G(T, P, L, N_1, N_2, \cdots, N_k) \#(1)$$

whose full differential is

$$dG = \left(\frac{\partial G}{\partial T}\right)_{P,L,N_j} dT + \left(\frac{\partial G}{\partial P}\right)_{T,L,N_j} dP + \left(\frac{\partial G}{\partial L}\right)_{P,T,N_j} dL + \sum_{i=1}^{k} \left(\frac{\partial G}{\partial N_i}\right)_{P,T,L,N_{j\neq i}} dN_i \#(2)$$

where $P$ is the pressure, $T$ is the temperature, $L$ is the adsorption displacement, $N_j$ is the number of particles of the j-th component in the adsorbate, and the corner mark $N_{j\neq i}$ indicates that the amount of the other components except the *i*-th component remains the same.

Because the adsorption work done by the system in a reversible process is equal to the decreasing value of the free energy, there is

$$\left(\frac{\partial G}{\partial L}\right)_{P,T,N_j} dL = f dL = -\delta W \#(3)$$

where $f = \left(\frac{\partial G}{\partial L}\right)_{P,T,N_j}$ is the generalized adsorption force, and $\left(\frac{\partial G}{\partial L}\right)_{P,T,N_j} dL$ is the contribution of the adsorption system at certain equipotential surface to the free energy change with respect to the adsorption displacement *L*.

In addition, the adsorption process is mostly carried out at constant temperature and pressure, i.e. $dT = dP = 0$, Eq. (2) can be simplified as

$$dG = -\delta W + \sum_{i=1}^{k} \left(\frac{\partial G}{\partial N_i}\right)_{P,T,L,N_{j \neq i}} dN_i \quad \#(4)$$

Since this adsorption process is spontaneous, according to the reduction principle of free enthalpy, i.e. $dG \leq 0$, there is

$$\sum_{i=1}^{k} \left(\frac{\partial G}{\partial N_i}\right)_{P,T,L,N_{j \neq i}} dN_i = \sum_{i=1}^{k} \mu_i dN_i \leq \delta W \#(5)$$

where the chemical potential $\mu_i = \left(\frac{\partial G}{\partial N_i}\right)_{P,T,L,N_{j \neq i}}$. Eq. (5) means the contribution of a multi-component system to the adsorption work when the amount of the substance changes under the condition of constant temperature and pressure.

When the amount $dN_i^b$ of the $i$-th component at the zero point of the bulk phase in the system is adsorbed to an equipotential surface of the adsorption phase, the amount added to the $i$-th component at the equipotential surface is $dN_i^a$, and $-dN_i^b = dN_i^a = dN_i$, thus Eq. (5) becomes

$$(\mu_i^a - \mu_i^b)dN_i \leq \delta W \#(6a)$$

which means the adsorption potential of an adsorbate molecule in the adsorption force field on the solid surface is the work required to move the molecule from its equipotential position in the adsorption phase to the zero position.

When $dN_i = 1$ mol, the work done by the system to adsorb 1 mol of substance is equal to the Polanyi molar adsorption potential ε, thus

$$A(\mu_i^a - \mu_i^b) \leq \varepsilon \#(6b)$$

where $A$ is Avogadro number, which denotes that ε is the change of the chemical potential when 1 mol of substance is reversibly adsorbed from the bulk phase $b$ to the adsorption phase $a$.

## 3. Adsorption condition of superconductors

For superconductors, this adsorption potential is also caused by the enhanced condensation, which should depend on the properties of the superconductor's lattice structure and the position of the electrons in the vicinity of the lattice surface. Since the concentration of electrons in the normal states is uniform, the adsorption potential is zero, that is, the interface between the normal states and the adsorption phase is zero, and the adsorption potential of the free electrons in equilibrium with the solid surface can be any equilibrium value from zero to saturation.

### 3.1 Thermodynamic adsorption condition for superconductors

According to Ref. [21], we have obtained the correlation among the electron concentration, temperature and pressure of Fermi electrons by the catastrophe theory as:

$$Pv^{5/3} = \left(\frac{\hbar^2}{m}\right)\left(\frac{mk_BT}{\hbar^2 n^{2/3}}\right)^{4\alpha/3} \cdot \left[\sqrt[3]{B + \sqrt{B^2 - \frac{1}{27}\left(\frac{mk_BT}{\hbar^2 n^{2/3}}\right)^\alpha}} + \sqrt[3]{B - \sqrt{B^2 - \frac{1}{27}\left(\frac{mk_BT}{\hbar^2 n^{2/3}}\right)^\alpha}}\right]^4 \quad \#(7)$$

where $\hbar$ is the Plank's constant, $m$ the rest mass of electron, $k_B$ is the Boltzmann constant, $n = N/V$ is the number of electrons per unit volume, $\alpha$ is the phase transition index, $\beta = \frac{mk_BT}{\hbar^2 n^{2/3}}$ is a dimensionless temperature, and $\beta^{1/2}$ shows a ratio of the mean distance among the particles $v^{1/3}$ and the thermal wavelength $\lambda = \sqrt{\frac{2\pi\hbar^2}{mk_BT}}$, and $B = \left[\frac{(3\pi^2)^{2/3}}{5}\right]^{1/4} + \left[\frac{(3\pi^2)^{2/3}}{5}\right]^{3/4}$.

From Eq. (7), when $\beta \gg 1$, which means the mean distance among the particles $v^{1/3}$ is much larger than the thermal wavelength $\lambda = \sqrt{\frac{2\pi\hbar^2}{mk_BT}}$, thus the microcosmic quantum effect can be ignored. In the infinite limit of $\beta \to \infty$ and $\alpha = 1/2$, Eq. (7) will degenerate to the state equation of the ideal Boltzmann gas with $P = nk_BT$. [21] When $\alpha = 0, \beta \to 0$, i.e. $\beta^\alpha = 1$, the pressure of the Fermi gas is not equal to zero because of the Pauli exclusion principle with $P = n^{5/3}\frac{\hbar^2}{m}\frac{(3\pi^2)^{2/3}}{5}$ [21].

For those high-temperature superconductors, such as Perovskite copper oxides (HgBa$_2$Ca$_2$Cu$_3$O$_{8+\delta}$), there is high superconductivity temperature (up to $T_c = 135$K) with low carrier concentration ($n = 8.85 \times 10^{27}$). And for those low-temperature superconductors, such as Ag, there is low superconductivity temperature (0.9K) with high carrier concentration ($n = 1.044 \times 10^{31}$). Anyway there is $\beta = \frac{mk_BT}{\hbar^2 n^{2/3}} \ll 1$ for all superconductors. Therefore, in this case Eq. (7) can be simplified as

$$P = 1.74 n k_B T \left(\frac{mk_BT}{\hbar^2 n^{2/3}}\right)^{4\alpha/3 - 1} \quad \#(8a)$$

i.e.,

$$n = \left(\frac{1.74\left(\frac{\hbar^2}{m}\right)\left(\frac{mk_BT}{\hbar^2}\right)^{4\alpha/3}}{P}\right)^{\frac{9}{8\alpha - 15}} \quad \#(8b)$$

Eq. (8) is a thermodynamic quantum equation of state of Fermi gases, which means that under the same pressure, when the electron concentration $n$ is increased, $T$ is decreased.

By use of Eq. (8a), further the Gibbs free energy of $N$ identical electrons is [21]

$$G(P,V,T) = F(P,V,T) + PV = Nk_BT\left[2.638\left(\frac{mk_BT}{\hbar^2 n^{2/3}}\right)^{\frac{5}{3}\alpha-1} + 12.88\left(\frac{mk_BT}{\hbar^2 n^{2/3}}\right)^{\frac{4}{3}\alpha-1}\right] \#(9)$$

where $F(P,V,T)$ is the free energy of the system.

From Eq. (9), the chemical potential at the normal phase is

$$\mu_n = \left.\frac{\partial G}{\partial N}\right|_{T,V} = 4.397k_BT\left[\left(1-\frac{2}{3}\alpha\right)\left(\frac{mk_BT}{\hbar^2 n^{2/3}}\right)^{\frac{5}{3}\alpha-1} + 4.88\left(1-\frac{8}{15}\alpha\right)\left(\frac{mk_BT}{\hbar^2 n^{2/3}}\right)^{\frac{4}{3}\alpha-1}\right] \#(10)$$

which shows that the chemical potential is dependent on the temperature $T$, the electron concentration $n$ and the phase transition index $\alpha$.

On the other hand, at the adsorption phase, due to the work done by the adsorption force, the electron concentration increases, resulting in a decrease in temperature $T$ according to Eq. (8a). With $T$ decreased to the superconductivity temperature $T_c$, some additional form of electron order begins to form and increases as the temperature drops. At this time the phase transition index $\alpha$ becomes $\alpha'$, and the superconducting electron concentration becomes $n_s = \omega n\ (\omega \leq 1)$, where $\omega$ is the relative proportion.

Because the special properties of the superconducting state are determined by the condensed matter of the Cooper pairs. The formation of Cooper pairs allows electrons to flow through the material without resistance, and this new electron pairing state plays a dominant role in the energy and particle number changes of the system. Therefore, in the superconducting state, the chemical potential is mainly considered about the superconducting electrons (Cooper pairs).

Then the chemical potential $\mu_s$ of the superconducting phase is

$$\mu_s = \left.\frac{\partial G}{\partial N_s}\right|_{T,V} = \frac{4.397k_BT}{\omega}\left[\left(1-\frac{2}{3}\alpha'\right)\left(\frac{mk_BT}{\hbar^2 n^{2/3}}\right)^{\frac{5}{3}\alpha'-1} + 4.88\left(1-\frac{8}{15}\alpha'\right)\left(\frac{mk_BT}{\hbar^2 n^{2/3}}\right)^{\frac{4}{3}\alpha'-1}\right] \#(11)$$

where $N_S$ is the number of superconducting electrons.

Therefore, according to Eq. (6b), the adsorption potential condition for the superconducting phase transition is

$$\varepsilon_p = \frac{\varepsilon}{A} \geq \mu_s - \mu_n \#(12)$$

where $\varepsilon_p$ is the average adsorption energy of every electron.

### 3.2 Identification of the phase transition index $\alpha$ for superconductors

When a normal metal is cooled, there is usually a decrease in conductive electron entropy. At temperatures less than $T_c$, some additional form of electronic ordering must begin to form, thus making an additional contribution to the specific heat capacity.

According to Eqs. (8) and (9), the conductive electron entropy $S$ and the specific heat capacity $C_p$ can be further obtained respectively as

$$S = \frac{\partial G}{\partial T}\bigg|_P = \frac{k_B \alpha}{8\alpha - 15}\left[44.8494\left(\frac{mk_B T}{\hbar^2 n^{2/3}}\right)^{5\alpha/3-1} + 154.56\left(\frac{mk_B T}{\hbar^2 n^{2/3}}\right)^{4\alpha/3-1}\right] \#(13)$$

and

$$C_p = T\frac{\partial S}{\partial T}\bigg|_p = \frac{15 k_B \alpha}{(8\alpha - 15)^2}\left[\begin{array}{l}44.8494\left(\frac{5\alpha}{3} - 1\right)\left(\frac{mk_B T}{\hbar^2 n^{2/3}}\right)^{5\alpha/3-1} + \\ 154.56\left(\frac{4\alpha}{3} - 1\right)\left(\frac{mk_B T}{\hbar^2 n^{2/3}}\right)^{4\alpha/3-1}\end{array}\right] \#(14)$$

From the point of view of entropy, the formation of Cooper pairs and their motion states affect the calculation of entropy. Therefore, when calculating the entropy of a superconducting state, the influence of superconducting electrons is mainly considered, i.e., the electron contribution $n$ in Eq. (13) will be replaced by the superconducting electron concentration $n_s = \omega n$.

For those low-$T_c$ superconductors, it is well known that $C_p \propto T^3$ at the superconductivity phase and $C_p \propto T$ at the normal phase. Considering Eq. (8b) where $n \sim T^{\frac{12\alpha}{8\alpha-15}}$ and $\beta = \frac{mk_B T}{\hbar^2 n^{2/3}} < 1$, from Eq. (13) we have $\alpha' = \frac{15}{11}$ for $T \leq T_c$ and $\alpha = \frac{15}{14}$ for $T > T_c$, which results in a jump in the specific heat capacity $C_p$ at $T_c$. Meanwhile, since the superconducting phase transition does not undergo latent heat changes, from the continuity of the entropy $S$ at $T_c$, the change of electron concentration from the normal phase to the superconductivity phase can also be uniquely determined, i.e. the relative proportion parameter $\omega$ of $n_s$ is identified.

In fact, when a superconducting phase transition occurs in a material, since electrons form Cooper pairs, their motion states are bound, and the interaction mode between the electrons changes, so that the effective interaction potential energy among the electrons changes, resulting in the pressure of the electron gas changing at the same time. Thus for the superconductivity phase, according to Eq. (8b) there is

$$P_s = 1.74 n_s k_B T_c \left(\frac{mk_B T_c}{\hbar^2 n_s^{2/3}}\right)^{4\alpha'/3-1} \#(15)$$

where $P_s$ is the pressure of the electron gas at the superconductivity phase.

The following is an example of metal $S_n$ to illustrate its superconducting phase transformation process. For the metal $S_n$, $T_c = 3.72$K, at which $n = 8.8 \times 10^{28}$. According to the continuity of the entropy $S$ at $T_c$, from Eq. (13) we can obtain that $\omega = 0.0006223$, which means that the phenomenon of superconductivity will occur only when the superconducting electron concentration is increased to $0.6223$ ‰ under the action of adsorption potential.

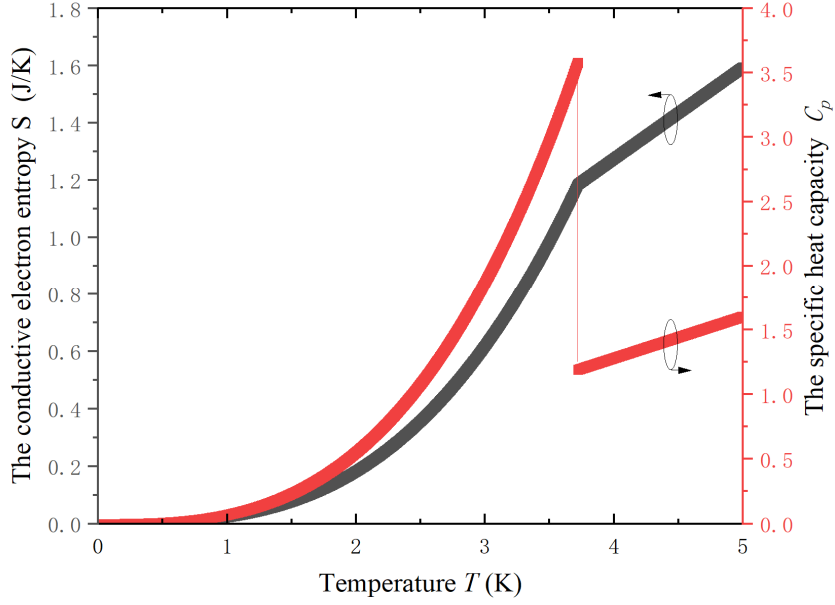

Fig.2 The entropy and the specific heat capacity of Sn varying with Temperature

Figure 2 shows the conductive electron entropy $S$ and the specific heat capacity $C_p$ of Sn of 1 molar electrons varying with temperature $T$ by using Eqs. (13) and (14).

### 3.3 The adsorption potential condition for superconductors

Based on the above section, the adsorption potential condition for the superconducting phase transition can be further obtained as

$$\varepsilon_p \geq \mu_s - \mu_n = k_B T \left[ \begin{array}{c} \dfrac{0.4}{\omega}\left(\dfrac{mk_BT}{\hbar^2 n^{2/3}}\right)^{\frac{14}{11}} + \dfrac{5.856}{\omega}\left(\dfrac{mk_BT}{\hbar^2 n^{2/3}}\right)^{\frac{9}{11}} \\ -1.256\left(\dfrac{mk_BT}{\hbar^2 n^{2/3}}\right)^{\frac{11}{14}} - 9.194\left(\dfrac{mk_BT}{\hbar^2 n^{2/3}}\right)^{\frac{3}{7}} \end{array} \right] \#(16a)$$

where $\omega$ is determined by the continuity of the entropy $S$ at $T_c$ from Eq. (13).

According to Eq. (16a), when $T \to 0$, $\omega \to 1$, and $\varepsilon_p \to 0$, which means that the adsorption potential is different from the BCS theory in which the band gap $2\Delta(0) = 3.53 k_B T_c$. On the other hand, when $T = T_c$, there is

$$\varepsilon_p \geq \mu_s - \mu_n = k_B T_c \left[ \begin{array}{c} \dfrac{0.4}{\omega}\left(\dfrac{mk_BT_c}{\hbar^2 n_c^{2/3}}\right)^{\frac{14}{11}} + \dfrac{5.856}{\omega}\left(\dfrac{mk_BT_c}{\hbar^2 n_c^{2/3}}\right)^{\frac{9}{11}} \\ -1.256\left(\dfrac{mk_BT_c}{\hbar^2 n_c^{2/3}}\right)^{\frac{11}{14}} - 9.194\left(\dfrac{mk_BT_c}{\hbar^2 n_c^{2/3}}\right)^{\frac{3}{7}} \end{array} \right] \#(16b)$$

which is also different from the BCS theory where there is the band gap $\Delta(T_c) = 0$, here $n_c$ is the electron concentration at $T_c$, and $n_s = \omega n_c$. Therefore, at $T=T_c$ the adsorption potential still works despite the disappearance of the energy gap of the

BCS theory.

Table 1 The average adsorption energy of every electron $\varepsilon_p$ for superconductors

| Super-conductors | The superconductivity temperature $T_c$ (K) | The electron concentration at $T_c$ | The relative proportion of superconducting electron concentration ω | The average adsorption energy of every electron $\varepsilon_p$ | The molar adsorption potential ε (J) |
|---|---|---|---|---|---|
| Ag | 0.9 | $3.26 \times 10^{30}$ | $2.898 \times 10^{-6}$ | $0.21 k_B T_c$ | 1.57 |
| Sn | 3.72 | $8.8 \times 10^{28}$ | 0.0006223 | $0.41 k_B T_c$ | 12.67 |
| $NbS_2$ [22] | 7.3 | $5.0 \times 10^{27}$ | 0.0039714 | $0.44 k_B T_c$ | 26.94 |
| $La_{2-x}Sr_xCuO_4$ [22] | 40 | $1.5 \times 10^{27}$ | 0.0238484 | $0.32 k_B T_c$ | 106.37 |
| $YBa_2Cu_3O_{6.7}$ [22] | 60 | $1.5 \times 10^{27}$ | 0.03186 | $0.2665 k_B T_c$ | 132.89 |
| $YBa_2Cu_3O_7$ [22] | 90 | $3.0 \times 10^{27}$ | 0.0305976 | $0.27485 k_B T_c$ | 205.54 |
| $HgBa_2Ca_2Cu_3O_{8+\delta}$ [23] | 135 | $8.85 \times 10^{27}$ | 0.0244192 | $0.316 k_B T_c$ | 354.59 |

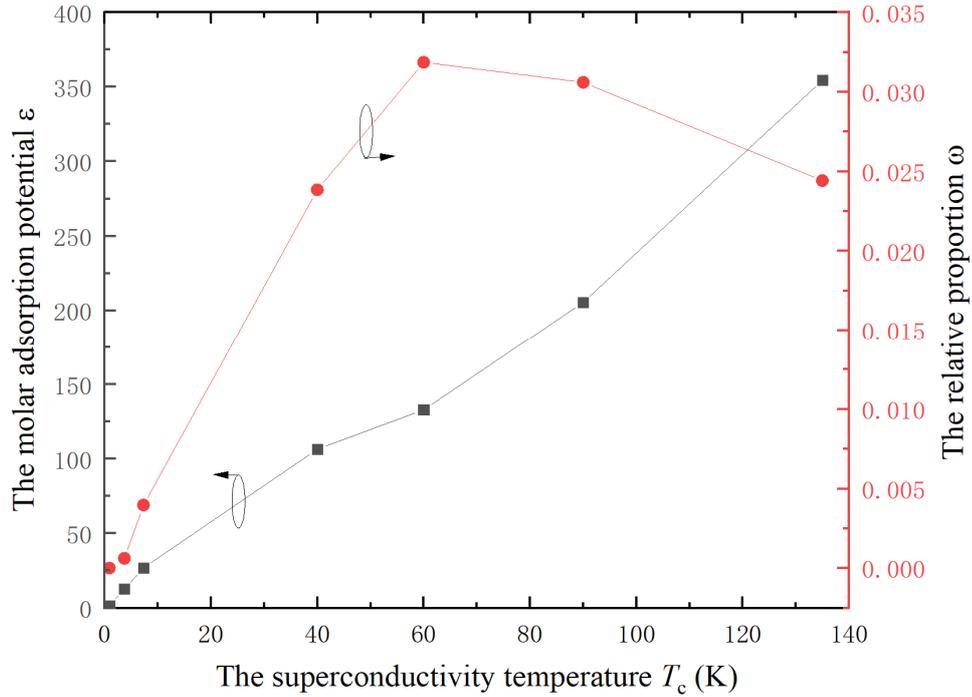

Fig.3 The molar adsorption potential ε and the relative proportion of superconducting electron concentration ω varying with the superconductivity temperature $T_c$

According to Eq. (16b), in Table 1 we calculate the average adsorption energy of every electron $\varepsilon_p$ and the corresponding molar adsorption potential ε for superconductors, respectively. Furthermore, the molar adsorption potential ε and the relative proportion of superconducting electron concentration ω varying with the superconductivity temperature $T_c$ are plotted in Fig. 3. It can be found that the molar adsorption potential ε is almost proportional to the superconductivity temperature $T_c$, and with higher slope when $T_c < 40K.$ These results reveal that those high-$T_c$ superconductors ($T_c \geq 40K$) are mainly formed by the molar adsorption potentials, and the low-$T_c$ superconductors ($T_c < 40K$) by both the molar adsorption potentials and the energy gap of the BCS theory. In addition, the relative proportion ω at $T_c$ in the high-$T_c$ superconductors is much more than that in the low-$T_c$ superconductors. Actually ω reflects the composition and structure of superconductors, which are of course decisive for the adsorption potentials, even though with the same electron concentration. Thus the following fact could be explained that for the Copper oxides with high-temperature superconducting, the smallest cell is at least an intact cell layer containing the CuO$_2$ bilayer.

For these low-$T_c$ superconductors with high carrier concentration at the normal phase, their composition and structure are so simple that the adsorption potential is too low to form additional Cooper-pairs apart from the electron-phonon interaction mechanism. On the other hand, for those high-$T_c$ superconductors but with low carrier concentration at the normal phase, when the degree of electron migration ω is increased to some extent, the molar adsorption energy is increased resulting in the high-$T_c$ phase transition. From this point of view, the anomaly of the superconductivity property could be explained for those high-temperature superconductors, and meanwhile the anomaly of the normal states could be explained for those low-temperature superconductors.

## 4. Conclusions

In this paper, the Molar adsorption potential is defined exactly, and the thermodynamic adsorption potential condition is put forward to explain the high-temperature superconductivity. It is revealed that beyond the electron-phonon interaction mechanism, the Cooper-paired electrons are mainly formed by this physical adsorption potential for high-$T_c$ superconductors. Thus the high-temperature superconductivity could still be explained by the Cooper-paired electrons due to the Molar adsorption potential. This theory could explain many anomalies of the normal and superconducting states of Perovskite copper oxides, as well as the isotopic effects of copper oxides.